\documentclass[english,twocolumn,prx,amssymb,amsmath,superscriptaddress]{revtex4}
\usepackage{graphicx,color}
\usepackage[bookmarks]{hyperref}
\usepackage{babel}
\usepackage{bbold}
\usepackage{braket}

\newcommand{\be}{\begin{equation}}
\newcommand{\ee}{\end{equation}}
\newcommand{\bea}{\begin{eqnarray}}
\newcommand{\eea}{\end{eqnarray}}

\definecolor{mygreen}{rgb}{0,0.5,0}
\definecolor{myblue}{rgb}{0,0,0.75}
\definecolor{mymagenta}{cmyk}{0,1,0,0.12}

\usepackage{umoline}

\newcommand{\sect}[1]{~\\\noindent{\large{\bf{#1}}~\\~\\ \noindent}}
\newcommand{\subsect}[1]{\noindent{\bf{#1}}\\ \noindent}

\newcommand{\ue}{\ensuremath{{\rm{e}}}}
\newcommand{\ud}{\ensuremath{{\rm{d}}}}

\begin{document}

\title{Quantum localization bounds Trotter errors in digital quantum simulation}
\author{Markus Heyl}
\email{heyl@pks.mpg.de}
\affiliation{Max Planck Institute for the Physics of Complex Systems, N\"othnitzer Str.38, 01187,Dresden, Germany}
\author{Philipp Hauke} 
\email{philipp.hauke@kip.uni-heidelberg.de}
\affiliation{Kirchhoff-Institute for Physics, Heidelberg University, 69120 Heidelberg, Germany}
\affiliation{Institute for Theoretical Physics, Heidelberg University, 69120 Heidelberg, Germany}
\author{Peter Zoller} 
\email{peter.zoller@uibk.ac.at}
\affiliation{Institute for Quantum Optics and Quantum Information of the Austrian Academy of Sciences, 6020 Innsbruck, Austria}
\affiliation{Institute for Theoretical Physics, University of Innsbruck, 6020 Innsbruck, Austria}

\date{\today}

\maketitle

{\bf
A fundamental challenge in digital quantum simulation (DQS) is the control of inherent errors. These appear when discretizing the time evolution generated by the Hamiltonian of a quantum many-body system as a sequence of quantum gates, called Trotterization. Here, we show that quantum localization--by constraining the time evolution through quantum interference--strongly bounds these errors for local observables. Consequently, for generic quantum many-body Hamiltonians, Trotter errors can become independent of system size and total simulation time. For local observables, DQS is thus intrinsically much more robust than what one might expect from known error bounds on the global many-body wave function. This robustness is characterized by a sharp threshold as a function of the Trotter step size. The threshold separates a regular region with controllable Trotter errors, where the system exhibits localization in the space of eigenstates of the time-evolution operator, from a quantum chaotic regime where the trajectory is quickly scrambled throughout the entire Hilbert space. Our findings show that DQS with comparatively large Trotter steps can retain controlled Trotter errors for local observables. It is thus possible to reduce the number of quantum gate operations required to represent the desired time evolution faithfully, thereby mitigating the effects of imperfect individual gate operations.}

\begin{figure}
	\centering
	\includegraphics[width=\columnwidth]{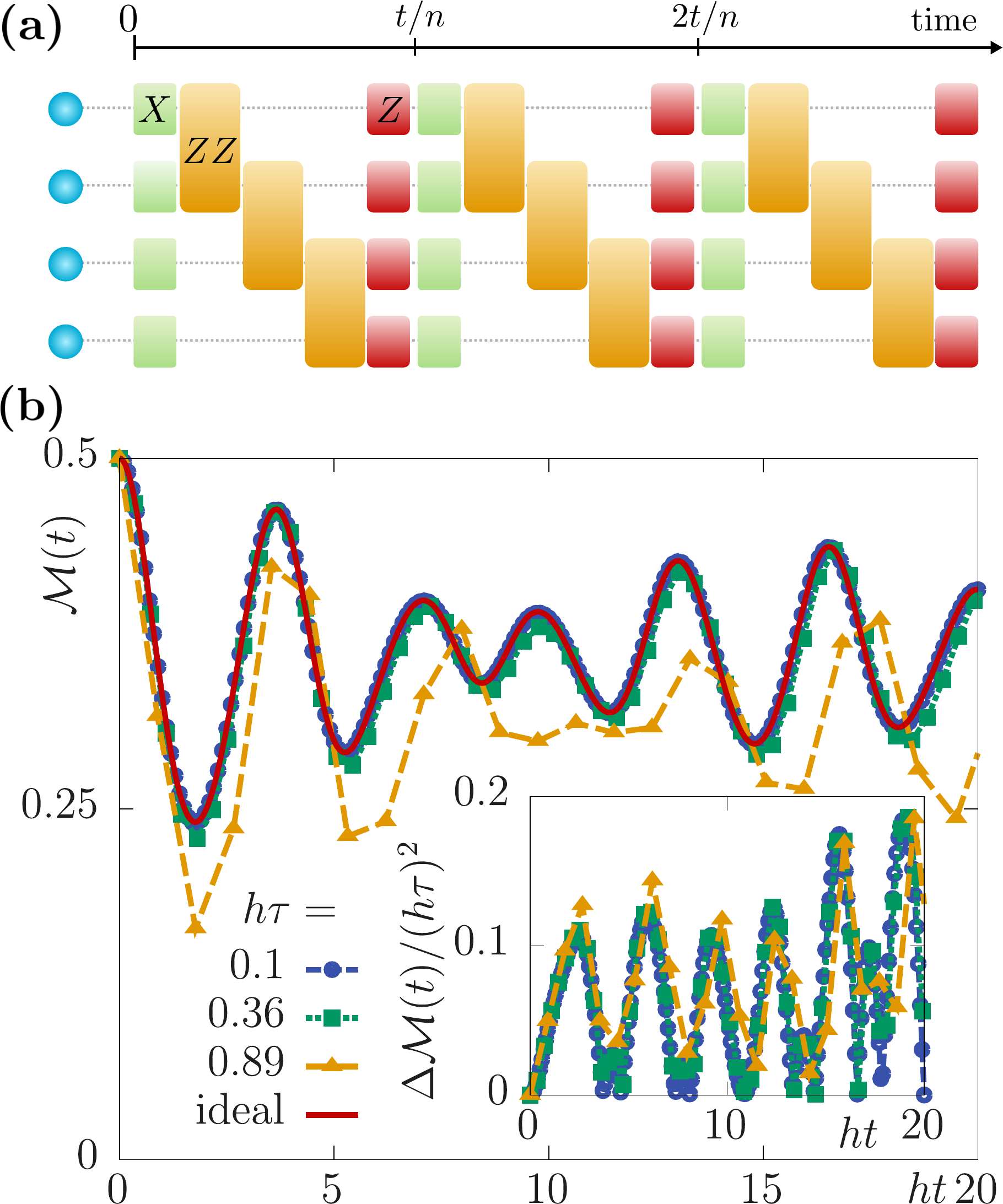}
	\caption{
		{\bf Trotterized time evolution and resulting error on local observables.}
		{\bf (a)} Gate sequence for the digital quantum simulation (DQS) of an Ising model. The desired evolution up to total simulation time $t$ is split into $n$ repeated sequences of length $\tau=t/n$, each decomposed into fundamental quantum gates. The example shows a gate sequence for a 4-qubit chain with Ising spin--spin interactions ($ZZ$) and transverse and longitudinal fields (simulated by single-qubit operations along the $X$ and $Z$ directions on the Bloch sphere). 
		{\bf (b)} Magnetization dynamics $\mathcal{M}(t) = N^{-1} \sum_{l=1}^N \langle S_l^z(t)\rangle$ in the DQS of the Ising model for $N=20$ spins and different Trotter step sizes $\tau$ compared to the exact solution. The normalized deviation $\Delta \mathcal{M}(t)/(h\tau)^2$ with $\Delta \mathcal{M}(t)=|\mathcal{M}_{\tau=0}(t)-\mathcal{M}(t)|$  from the ideal dynamics $\mathcal{M}_{\tau=0}(t)$ shows a collapse of the error dynamics for sufficiently small $\tau$.
	} 
	\label{fig:1}
\end{figure}

\sect{Introduction}
Quantum computers promise to solve certain computational problems exponentially faster than any classical machine~\cite{Ladd2000}. 
A particularly promising application is the solution of quantum many-body problems \cite{Feynman1982}, with large potential impact on quantum chemistry, material science, and fundamental physics. 
The devices employed in this effort can be divided into two major classes: analog quantum simulators, where the  Hamiltonian of interest is engineered to mimic the desired quantum many-body physics; and digital quantum simulators (DQSs), where a target time-evolution operator is represented by a sequence of elementary quantum gates.
The digital approach is particularly flexible, since a universal DQS can be freely programmed to simulate the unitary evolution of any many-body Hamiltonian with local interactions~\cite{Lloyd1996} (Fig.~\ref{fig:1}a). 
Recent experiments have demonstrated remarkable progress in implementing DQS, e.g., by simulating simple molecules in quantum chemistry \cite{OMalley2016,Kandala2017,hempel2018quantum}, condensed-matter models \cite{Lanyon2011,Salathe2015,Barreiro2011,Barends2015,Langford2016,Wei2018}, and lattice gauge theories \cite{Martinez2016}. 
The working principle of DQS is as follows. Suppose that the target Hamiltonian $H=\sum_{l=1}^M H_l$ can be decomposed into $M$ terms whose time evolution operators $U_l(t) = \exp(-it H_l)$ can be implemented on the considered quantum computing device.
Using the Suzuki--Trotter formula, the full time-evolution operator $U(t) = \exp(-itH)$ can be approximated by discretizing it into $n \in \mathbb{N}$ repetitions of the fundamental gates $U_l$: 
\be
	U^{(n)}(t) = \left[ U_1\left(\frac{t}{n}\right) U_2\left(\frac{t}{n}\right) \cdots U_M\left(\frac{t}{n}\right) \right]^n\,.
	\label{eq:Trotterization}
\ee
This Trotterization comes inherently with an error that can be rigorously bounded via the accuracy of the global unitary time-evolution operator~\cite{Lloyd1996}
\be
	U(t) - U^{(n)}(t) = \frac{t^2}{2n} \sum_{l>m=1}^M  [H_l,H_m] + \epsilon\,.
	\label{eq:LloydErrorBound}
\ee
Here, $\epsilon$ subsumes terms of order $t^3/n^2$ and higher.
Consequently, for the lowest order corrections the error grows quadratically with total simulation time $t$ and (in generic quantum many-body systems) linearly in the number of simulated degrees of freedom $N$.  
It is possible to improve this bound, but an error bound that scales less than linear in $t$ is not possible if one is concerned with the entire unitary operator~\cite{Berry2007}. 
Although the polynomial scaling with both $t$ and $N$ is efficient in a computational complexity sense, it poses a significant challenge for practical computations~\cite{poulin2014trotter,Babbush2015}, seemingly preventing current technology from simulating all but small instances. 
As we show in this article, these generic bounds on the global many-body wave function overestimate by far the actual error on local observables such as magnetizations or low-order correlation functions.
For example, in the DQS of a quantum Ising chain the deviation of the magnetization dynamics from the ideal evolution can be significantly smaller and remain bounded even at long times, see Fig.~\ref{fig:1}b and inset.
It is the purpose of this article, to explain this observation from physical grounds, and thus assign a physical interpretation to Trotter errors.

We achieve this by linking Trotterization errors to quantum localization.
Localization is a ubiquitous phenomenon with many facets.
Initially, it has been introduced to understand the absence of transport in systems of free particles with disorder~\cite{Anderson1958}.
Since then, the concept has been generalized to various contexts such as many-body localization in Hilbert space as absence of quantum ergodicity~\cite{BASKO2006} or energy localization in periodic time-dependent quantum many-body systems as absence of heating in continuously driven systems~\cite{DAlessio2013}.
As we show here, at small Trotter steps a related  localization in Hilbert space occurs that bounds time-discretization errors on local observables.

\sect{Results}
\subsect{Trotter sequences as Floquet systems}
In this work, we interpret the Trotterized evolution as a periodically time-dependent quantum many-body system with a period $\tau=t/n$, see Fig.~\ref{fig:1}.
The desired stroboscopic dynamics is therefore governed by an associated Floquet Hamiltonian $H_F$, which we define for later convenience in the following form:
\be
  e^{-iH_F\tau} = U_1\left(\tau\right) U_2\left(\tau\right) \cdots U_M\left(\tau\right) \, ,
  \label{eq:defFloquetHamiltonian}
\ee
The starting point of our considerations is an analytical expression for $H_F$ in the limit of sufficiently small Trotter steps $\tau$, 
\be
{H}_F = H + i\frac{\tau}{2} \sum_{l>m} [H_l,H_m] + \mathcal{O}(\tau^2). 
\label{eq:FloquetExpansion}
\ee
This form, which can be obtained from Eq.~(\ref{eq:defFloquetHamiltonian}) via a Magnus expansion, quantifies the Trotterization error on a Hamiltonian level.
There remain, however, two fundamental questions that we aim to address in this work: 
(i) What is the radius of convergence $\tau^\ast$ of this expansion? (ii) What is the influence of corrections to $H$ that appear in $H_F$ on the long-time dynamics of observables? 
Recent theoretical predictions for heating in generic quantum many-body systems subject to a periodic drive might leave a rather pessimistic impression \cite{Lazarides2014,Abanin2015,Mori2016}. We show in this work that the errors on local observables can nevertheless be controlled for all practical purposes.
\\

\subsect{Benchmark model: quantum Ising chain}
\label{sec:models}
In the following, we illustrate our discussion with a generic, experimentally relevant model, the quantum Ising chain with Hamiltonian $H = H_Z + H_X$, with $H_Z = J\sum_{l=1}^{N-1} S_l^z S_{l+1}^z + h\sum_{l=1}^N S_l^z$ and $H_X= g \sum_{l=1}^N S_l^x$. Here, $S_l^\gamma$, $\gamma=x,y,z$, denote spin-$1/2$ operators at lattice sites $l=1,\dots,N$. Such models are paradigmatic workhorses for DQS platforms such as nuclear magnetic resonance \cite{Peng2005}, trapped ions \cite{Lanyon2011}, and superconducting qubits~\cite{Barends2016}. 
As initial state, we choose $\ket{\psi_0} = \bigotimes_l \ket{\uparrow}_l$, which can be prepared with high fidelity~\cite{Barends2016,Lanyon2011,Jurcevic2016}.
In the remainder, we use the parameters $h/J=g/J=1$. 
For details about the simulations including the used gate sequences, see Methods.
Though we focus on this model, our findings also apply to various other model systems, and thus seem generic~\cite{Sieberer2018}, see also the Supplementary Materials where we provide a similar analysis for the lattice Schwinger model.\\  

\begin{figure}
	\centering
	\includegraphics[width=\columnwidth]{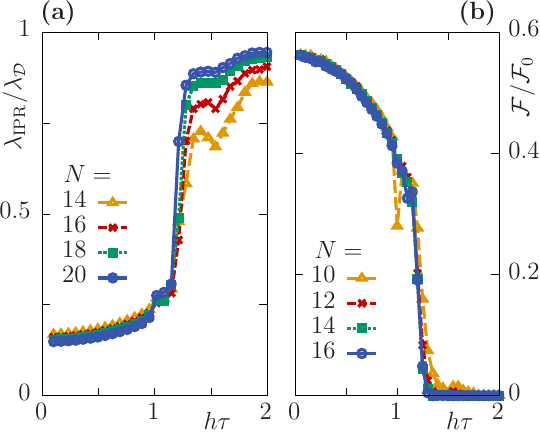}
	\caption{{\bf Localization and quantum chaos in the Trotterized dynamics of the quantum Ising chain.} {\bf(a)} Rate function $\lambda_\mathrm{IPR}$ of the inverse participation ratio, normalized to the maximally achievable value $\lambda_\mathcal{D}$ describing uniform delocalization over all accessible states. A sharp threshold as a function of the Trotter step size $\tau$ separates a localized regime at small $\tau$ from a quantum chaotic regime at large $\tau$. {\bf(b)} The long-time limit $\mathcal{F}$ of the out-of-time ordered correlator also signals a sharp quantum chaos threshold. $\mathcal{F}$ is normalized with respect to $\mathcal{F}_0=1/8$, the theoretical maximum. Full scrambling is only achieved for large Trotter steps.}
	\label{fig:chaos}
\end{figure}

\subsect{Quantum many-body chaos threshold}
As the central result of this work, we connect Trotter errors in DQS with a threshold separating a many-body quantum chaotic region from a localized regime, thus linking the intrinsic accuracy of a DQS with a quantum many-body phenomenon.
For that purpose, we first investigate the inverse participation ratio
\be
	\mathrm{IPR} = \sum_\nu p_\nu^2, \quad p_\nu = | \langle \phi_\nu | \psi_0\rangle|^2 \, ,
\ee
with $|\phi_\nu\rangle$ denoting a full set of eigenstates of the Floquet Hamiltonian $H_F$. 
The $\mathrm{IPR}$ measures the localization properties of the state $|\psi_0\rangle$ in the eigenbasis $|\phi_\nu\rangle$, which is well studied also in the single-particle context~\cite{Haake2010}.
In a quantum chaotic delocalized regime, $|\psi_0\rangle$ is scrambled across the full eigenbasis implying a uniform distribution $p_\nu \to \mathcal{D}^{-1}$, with $\mathcal{D}$ the number of available states in Hilbert space.
Since $\mathcal{D}$ grows exponentially with the number of degrees of freedom $N$, we introduce the rate function $\lambda_\mathcal{D} = N^{-1} \log( \mathcal{D} )$, which exhibits a well-defined thermodynamic limit.
Analogously, we define $\lambda_\mathrm{IPR} = -N^{-1} \log(\mathrm{IPR})$.
In Fig.~\ref{fig:chaos}a, we show numerical data for the ratio $\lambda_\mathrm{IPR}/\lambda_\mathcal{D}$ for the considered benchmark example.
For the data in this plot, we take into account the expected leading-order finite-size corrections $\lambda_\mathcal{D}=N^{-1}[\log(\mathcal{D})-\log(2)]$ in the delocalized regime, which can be estimated using random matrix theory~\cite{Ullah1963}.
As one can see, there appears a sharp threshold separating a quantum chaotic regime at large Trotter steps, where $\lambda_\mathrm{IPR}$ tends to $\lambda_\mathcal{D}$ with increasing system size, from a regular region with $\lambda_\mathrm{IPR}/\lambda_\mathcal{D} < 1$.
A strong fingerprint of quantum chaos can also be found in out-of-time ordered (OTO) correlators, which quantify how fast quantum information scrambles through a many-body system. A typical OTO correlator is of the form 
\be
    \mathcal{F}(t) = \langle V^\dagger(t) \, W^\dagger \, V(t) \, W \rangle\, ,
	\label{eq:defOTOC}
\ee
where $V(t)$ denotes the time evolution of the operator $V$ in the Heisenberg picture.
While quantum chaos via OTO correlators is conventionally diagnosed by considering a late-time exponential growth for operators $V$ and $W$ with finite support in real space~\cite{Maldacena2016}, here we consider the asymptotic long-time value of the extensive operator $V=W=N^{-1}\sum_l S_l^z$~\cite{Kukuljan2017}.
We estimate the corresponding long-time limit, $\mathcal{F} = \mathcal{F}(t\to\infty)$, via a stroboscopic average $\mathcal{F} = \lim_{n\to\infty} n^{-1} \sum_{l=1}^n  \, \mathcal{F}(l\tau)$.

In Fig.~\ref{fig:chaos}b, we present numerical evidence that this quantity detects the many-body quantum chaos threshold that we have seen in the IPR.
There is a clear threshold that separates a localized region at small Trotter steps $\tau$, where $\mathcal{F}>0$, from a quantum chaotic region at large $\tau$, where $\mathcal{F}\to 0$.
The vanishing OTO correlator in the many-body quantum chaotic regime can be understood directly from the results obtained for the IPR. 
Consider the spectral decomposition of a local Hermitian operator $V=\sum_\alpha \lambda_\alpha |\alpha \rangle \langle \alpha|$, with $\lambda_\alpha$ the eigenvalues and $|\alpha\rangle$ the eigenvectors of $V$ (for the considered magnetization, these are equivalent to the set of spin configurations).
The effective Floquet dynamics yields after $n$ periods
\be
  V(n\tau) = \sum_\alpha \sum_{\nu,\mu} \lambda_\alpha C_{\nu \alpha} C_{\mu \alpha}^\ast e^{-i(E_\nu-E_\mu)n\tau} | \phi_\nu\rangle \langle \phi_\mu| \, ,
\ee
with $E_\nu$ the Floquet quasi-energy corresponding to the eigenstate $|\phi_\nu\rangle$ and $C_{\nu\alpha} = \langle \phi_\nu | \alpha\rangle$.
The behavior of the IPR suggests that for all spin configurations $p_{\nu \alpha} \equiv |C_{\nu \alpha}|^2 = \mathcal{D}^{-1}$ is uniformly distributed, such that the amplitudes $C_{\nu\alpha}$ are almost structureless and contain only a phase information, $C_{\nu \alpha} = \mathcal{D}^{-1/2} e^{i \varphi_{\nu \alpha}}$. 
After sufficiently many Floquet cycles, this phase information is randomized and scrambled by the unitary evolution, except when $\nu=\mu$, projecting the operator to the so-called diagonal ensemble~\cite{ETHReview2016}. Thus, for $n\to\infty$ one obtains $V(n\tau) \to \mathcal{D}^{-1} \sum_\alpha \lambda_\alpha \mathbb{1} $.
Here, $\mathcal{D}^{-1} \sum_\alpha \lambda_\alpha = \mathcal{D}^{-1} \mathrm{Tr} \, V$ is equivalent to the infinite-temperature average, which yields a vanishing value for the considered total magnetization.
In other words, the operator becomes completely scrambled over the full Hilbert space.

Within the localized phase, the amplitudes $C_{\nu \alpha}$ contain more structure than only the phase information, which yields a nonzero value for the OTO correlator. 
For small systems, such as for $N=10$ in Fig.~\ref{fig:chaos}b, one can observe additional structures in the crossover region, which vanish for larger $N$.
We attribute these to individual quantum many-body resonances, which can be resolved in small systems, but which merge for large $N$.
\\

\subsect{Robustness of local observables}
While the corrections due to time discretization are weak on a Hamiltonian level, as seen in the Magnus expansion in Eq.~(\ref{eq:FloquetExpansion}), there is \emph{a priori} no guarantee that the long-time dynamics is equally well reproduced. 
It is, e.g., well known for classical chaotic systems that even weak perturbations can grow quickly in time.
Here, we provide numerical evidence that in the localized regime the dynamics of observables remains constrained and controlled, even in the long-time limit.

\begin{figure}
	\centering
	\includegraphics[width=\columnwidth]{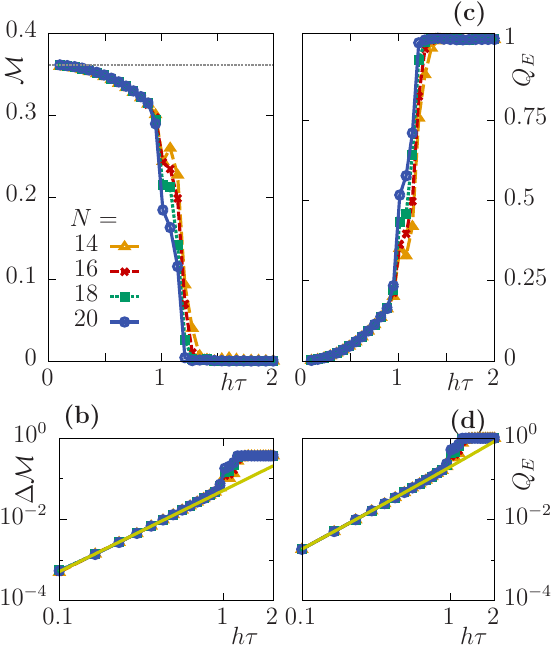}
	\caption{{\bf Trotter errors for local observables in the infinite long-time limit for the Ising model.}
	Both the magnetization $\mathcal{M}$ {\bf (a)} and simulation accuracy $Q_E$ {\bf (c)} exhibit a sharp crossover from a regime of controllable Trotter errors for small Trotter steps $\tau$ to a regime of strong heating at larger $\tau$. The dashed line in (a) refers to the desired case of the ideal evolution $\mathcal{M}_{\tau=0}$. The Trotter error exhibits a quadratic scaling at small $\tau$ for both the deviation of the magnetization, $\Delta \mathcal{M} = \mathcal{M} - \mathcal{M}_{\tau=0}$, {\bf (b)} and $Q_E$ {\bf (d)}. The solid lines in {\bf (b)} and {\bf (d)} represent analytical results obtained perturbatively in the limit of small Trotter steps $\tau$. These results indicate the controlled robustness of digital quantum simulation against Trotter errors, in the long-time limit and largely independent of $N$. }
	\label{fig:figRobustness}
\end{figure}

In Fig.~\ref{fig:figRobustness}a, we show the asymptotic long-time value $\mathcal{M}$ of the magnetization, $\hat{\mathcal{M}}(t) = N^{-1} \sum_{l} S_l^z(t)$.
One can clearly observe that the many-body quantum chaos threshold identified in the IPR and OTO correlator has a substantial influence on the long-time Trotter error of observables such as ${\mathcal{M}}(t)$.
For large Trotter steps $\tau$, the magnetization acquires its infinite-temperature value, perfectly consistent with the above analysis of the fully delocalized quantum chaotic phase.
Remarkably, however, for small Trotter steps the error $\Delta \mathcal{M}$ relative to the targeted dynamics exhibits a quadratic dependence in $\tau$, as we show in Fig.~\ref{fig:figRobustness}b.
The origin of these weak Trotter errors can already be identified from the dynamical trajectories of the magnetization shown in the inset of Fig.~\ref{fig:1}b, where we plot the error $\Delta \mathcal{M}(t)$ for different Trotter steps normalized with respect to $(h\tau)^2$.
We observe a collapse of trajectories corresponding to different $\tau$, with the overall magnitude of the error remaining bounded in time.
This finding suggests that in the localized phase the discretization error on observables itself behaves regular, in the sense that different perturbation strengths as measured by $\tau$ do not yield fast diverging expectation values.
\\

\subsect{Simulation accuracy\label{sec:simulationAccuracy}}
In the previous sections, we have provided evidence for a sharp threshold between a delocalized and a localized regime with controllable Trotter errors. 
We now aim to understand the influence of the regular regions onto the dynamics of observables.
We identify as the underlying reason for the weak Trotter errors a dynamical constraint due to an \emph{emergent} stroboscopic constant of motion in the effective time-periodic problem, which is the Floquet Hamiltonian $H_F$.
Although this integral of motion is different from the desired energy conservation of the target Hamiltonian $H$, the perturbative expansion in Eq.~(\ref{eq:FloquetExpansion}) suggests a close connection.
It is therefore natural to quantify the accuracy of a DQS by measuring how far the system deviates from the desired constant of motion $H$ via
\be
  Q_E(n\tau) \equiv \frac{E_{\tau}(n\tau) - E_0}{E_{T=\infty} - E_0}\,.
\ee
Here, we have introduced $E_\tau(n\tau)=\langle H(n\tau) \rangle_\tau$ and $E_0 = E_{\tau\to0}(n\tau) = \bra{\psi_0} H \ket{\psi_0}$, where the subindex $\tau$ refers to the used Trotter step for the dynamics.
In $Q_E(t)$, we normalize the errors using the system's energy at infinite-temperature, $E_{T=\infty} = \mathcal{D}^{-1} \mathrm{Tr} \, H$. 
In the idealized limit $\tau \to 0$, where the integral of motion $H_F \to H$, one has $Q_E(t)=0$.
In the opposite limit of large Trotter steps, i.e., in the many-body quantum chaotic region, we expect full delocalization over all eigenstates, yielding $Q_E(t)\to 1$ in the long-time limit.
Thus, $Q_E(t)$ defines a system-independent measure for the simulation accuracy.
From an alternative perspective, $Q_E(t)$ quantifies heating in the effective periodically driven system, as it has been studied previously in the context of energy localization~\cite{DAlessio2013}.
In Fig.~\ref{fig:figRobustness}c, we show numerical data for the long-time average $Q_E$.
Again, we find a sharp threshold between the localized and quantum chaotic regimes.
Importantly, for small Trotter steps $Q_E$ acquires only a weak quadratic dependence on $\tau$, see Fig.~\ref{fig:figRobustness}d, yielding
\be 
      Q_E \equiv Q_E(t\to\infty)  = (\tau/\tau_E)^{\alpha},\quad \tau \ll \tau_E, 
      \label{eq:simulationFidelityError}
\ee 
with $\alpha=2$.
While $\tau_E$ depends on the microscopic details of the system, we find from our numerics that there is no notable dependence on $N$ even in the asymptotic long-time limit, with potential corrections in the thermodynamic limit $N\to\infty$ that are discussed further below.

To obtain an analytical understanding for the observations of weak Trotter errors on local observables, let us start by considering the Magnus expansion for the Floquet Hamiltonian in Eq.~(\ref{eq:FloquetExpansion}), which quantifies the leading-order corrections due to time discretization on a Hamiltonian level.
From our numerical results for $Q_E$, we anticipate that the target Hamiltonian $H$ is an almost conserved quantity, which motivates us to study the perturbative corrections to strict energy conservation.
Using time-dependent perturbation theory up to second order in the Trotter step size $\tau$, we find
\be
	Q_E = q_E \, (h\tau)^2 + \mathcal{O} \left[ (h\tau^3) \right].
	\label{eq:QEtauanalytic}
\ee
The explicit derivation and the final formula for $q_E$ are given in the Methods.
For the considered parameters, we estimate $q_E=0.18$. As it can be seen in Fig.~\ref{fig:figRobustness}d, this analytical value matches well the numerical results. 

To test whether the errors on other local observables are also controlled by the emergent constant of motion in the localized regime, we exemplarily study the corrections to the targeted magnetization dynamics. 
From time-dependent perturbation theory, we obtain $\Delta M = m (J\tau)^2+\mathcal{O} \left[ \,(J\tau^3) \right]$ with $m=0.05$. This theoretical prediction is again very close to the numerical data (see Fig.~\ref{fig:figRobustness}b).
As these findings indicate, in the regular region at small Trotter steps the discretization error on local observables can be captured by time-dependent perturbation theory in the Trotter step size $\tau$ --- even in the asymptotic long-time limit.

Our observations give a smaller error on local observables than suggested by general considerations on Floquet dynamics in high-frequency regimes (corresponding to small Trotter steps)~\cite{KUWAHARA2016,Abanin2017}.
In these works, it is shown that there exists always a static local Hamiltonian $\tilde{H}$, different from $H$, which approximates the stroboscopic Floquet long-time dynamics. 
Our results show that the evolution of local observables is approximated by $H$ itself, as desired within DQS.

\sect{Discussion\label{sec:discussion}}
As we have shown, intrinsic Trotter errors in DQS are controllable for local observables, with a sharp threshold separating a localized from a many-body quantum chaotic regime.
While we show data here for one specific model, we observe similar behavior also for other generic systems with sufficiently short-ranged interactions~\cite{Sieberer2018}. In addition, also certain systems with long-range interactions can exhibit controllable Trotter errors~\cite{Sieberer2018} including also the recently experimentally realized Lattice Schwinger Model as we discuss in the Supplementary Materials.
Our numerical studies are based on up to $N=20$ qubits, which is within realized and expected size ranges of digital quantum simulators~\cite{OMalley2016,Kandala2017,hempel2018quantum,Lanyon2011,Salathe2015,Barreiro2011,Barends2015,Langford2016,Wei2018,Martinez2016, bernien2017, Zhang2017,Rigetti2017,Neill2018}. 
For experiments, it is of particular interest to assess the precise value of the threshold scale $\tau^\ast$.
Theoretically predicting $\tau^\ast$ is in general as difficult as solving the desired time-evolution. Nevertheless, one can estimate $\tau^\ast$ as follows. 
Before running an experiment, one can numerically calculate $Q_E$ for small $N$, yielding a first estimate on $\tau^\ast$.
From this starting point, experiments can find an optimal Trotter step at larger $N$ by decreasing $\tau$ until sufficient convergence is reached.
Once in the perturbative regime, one can use data at non-zero $\tau$ to extrapolate to the ideal dynamics in a well-defined way. 
While our results appear to be robust upon increasing the number of degrees of freedom, a quantitative extrapolation to $N\to\infty$ would require the numerical study of larger systems.
In this context, recent works have argued that in the thermodynamic limit generic periodically driven systems will eventually heat up indefinitely~\cite{Lazarides2014,Abanin2015,Mori2016}. 
This might leave a rather pessimistic impression, but, as we explain now, time discretization errors still remain controllable.
Even in the worst-case scenario where such an indefinite heating takes place, the energy growth can still be bounded on general grounds via $|E(t) - E_0| \leq Ce^{-\tau_0/\tau} t $ for $ \tau \ll \tau_0$~\cite{Abanin2015,Mori2016,Abanin2017,KUWAHARA2016}. 
Here, $C$ denotes a constant of dimension energy squared and $\tau_0$ a constant of dimension time, both of which are independent of $N$.
Thus, for a given total simulation time $t$, one can ensure a maximum allowed error $\Delta$ on the simulation accuracy $Q_E(t)$ by choosing $\tau$ according to $\tau = \tau_0/\log(ct/\Delta)$ with $c=C|E_{T=\infty}-E_0|$. 
In this worst-case scenario, the Trotter step size to reach a given accuracy therefore acquires at most a logarithmic dependence on $t$ but remains independent of $N$. This is still an exponential improvement over the global wave function bounds such as given in Eq.~(\ref{eq:LloydErrorBound}).
In practice, since it is tunable via $\tau$, this extremely slow intrinsic heating can always be adjusted such that the associated heating rate is smaller than that of other error sources, such that Trotter errors become insignificant.

Therefore, the accuracy of DQS experiments on local observables is limited mainly by extrinsic error sources. 
While these may in the future be eliminated by error correction \cite{Schindler2011,Reed2012}, for relevant system sizes to solve many-body problems full error correction is still out of reach with currently available resources. 
In the Supplementary Materials, we discuss in detail two typical extrinsic error sources, timing errors on individual gates and slow drifts of gate couplings over various shots of the experiment. 
The slow drifts turn out to be relatively benign, leading only to an effective average over an ensemble of target Hamiltonians. Individual timing errors, however, induce in the limit of small $\tau$ a time scale beyond which the accuracy of the DQS is severely affected. 
In addition, a realistic system will suffer from qubit decoherence as well as faulty pulses such as imperfect swaps between internal levels. Both of these make it highly preferable to use as few gates as possible. 
In view of these, our results become particularly relevant: as they show, intrinsic errors in a DQS remain controlled even with relatively large Trotter steps.
This makes it possible to reach a desired simulation time with a reduced number of gates, thus diminishing the influence of extrinsic errors and enhancing the accuracy of the DQS for local observables.


\subsect{Acknowledgments:}
We thank Anatoli Polkovnikov for invaluable discussions in the initial stages of this work and Lukas Sieberer, Andreas Elben, as well as Tobias Olsacher for various comments and suggestions on the manuscript.

\subsect{Funding:}
Work at Innsbruck was supported by ERC Synergy Grant UQUAM and the SFB FoQuS (FWF Project No. F4016-N23). 
M.~H. acknowledges support by the Deutsche Forschungsgemeinschaft via the Gottfried  Wilhelm  Leibniz  Prize  program.
P.~H. acknowledges support by the DFG Collaborative Research  Centre  SFB  1225  (ISOQUANT) and the ERC Advanced  Grant  EntangleGen (Project-ID  694561). 

\subsect{Author contributions:}
The project has been devised jointly by all the authors. All authors discussed the results and wrote the manuscript. The numerical simulations have been performed by M.~H.

\subsect{Competing interests:}
The authors declare no competing financial interests.

\subsect{Data and materials availability:}
All data needed to evaluate the conclusions in the paper are present in the paper and/or the Supplementary Materials. Additional data related to this paper may be requested from the authors.


\vspace{1cm}
\noindent{{\Large\bf Materials and Methods}}\\

\subsect{Numerical methods and gate sequences}
In the main text, we show numerical data for a quantum Ising chain with the Hamiltonian 
\be
H = H_Z + H_X\,,
\ee
where
\be
H_Z = J\sum_{l=1}^{N-1} S_l^z S_{l+1}^z + h\sum_{l=1}^N S_l^z, \quad H_X= g \sum_{l=1}^N S_l^x.
\ee
Many of the involved contributions in this model Hamiltonian mutually commute.
Therefore, only a small set of elementary quantum gates is required to simulate the Trotterized dynamics.
We use the following sequence of two gates:
\be
U^{(1)} = U_1 U_2,\quad U_1 = e^{-i\tau H_Z}, \,\, U_2= e^{-i\tau H_X}.
\ee
For the presented simulations of observables, we have computed the real-time evolution for $2 \cdot 10^4$ periods, except otherwise noted, using a Lanczos algorithm with full reorthogonalization. Because for a finite-size system observables still show remaining temporal fluctuations, we extract the asymptotic long-time limit of the presented quantities by performing a stroboscopic time average over the last $10^4$ periods. 

The inverse participation ratio shown in Fig.~\ref{fig:chaos} can, in principle, be obtained either by exact diagonalization or by use of a dynamical evolution. We have chosen the latter because it allows us to reach larger systems and is, in principle, an experimentally accessible approach. Dynamically, the inverse participation ratio can be obtained by a stroboscopic mean, 
\be
\mathrm{IPR} = \lim_{n\to\infty} \frac{1}{n} \sum_{l=1}^n \mathcal{P}_l, \quad \mathcal{P}_l = \Big| \langle \psi_0 | e^{-ilH_F\tau} | \psi_0 \rangle \Big|^2\,,
\ee
as one can prove by expanding $\mathcal{P}_l$ in the eigenbasis of $H_F$, followed by a summation of the resulting geometric series. 
Note that $\mathcal{P}_l$ is nothing else than the Loschmidt echo, a common indicator for quantum chaotic behavior in single-particle systems \cite{Haake2010}.\\

For the computation of the OTO correlator $\mathcal{F}(t)$ defined in Eq.~(\ref{eq:defOTOC}), 
we have decomposed $\mathcal{F}(n\tau)$ as
\be
	\mathcal{F}(n\tau) = \langle \psi_1(n\tau) |\psi_2(n\tau)\rangle \, ,
\ee
where the two states
\begin{align}
	|\psi_1(n\tau)\rangle = W e^{i H_F n\tau} V e^{-iH_Fn\tau}|\psi_0\rangle \, , \\
	|\psi_2(n\tau)\rangle = e^{i H_F n\tau} V e^{-iH_Fn\tau} W|\psi_0\rangle \, ,
\end{align}
can be obtained from forward and backward evolving the quantum many-body state with appropriate insertions of the $W$ and $V$ operators.
Since the backward evolution has to be performed for every Trotter step $n$, the overall runtime of this approach scales proportional to $n^2$.
This limits the accessible total simulation time $t=n\tau$, such that we have used $n=10^3$ for the simulations shown in the main text, and we have performed a stroboscopic average over the last $300$ periods to obtain an estimate for the asymptotic long-time value.\\

\subsect{Trotter errors on local observables from perturbation theory}
As mentioned in the main text, the Trotter errors for local observables can be captured using time-dependent perturbation theory in the limit of sufficiently small $\tau$.
In the following, we outline how to obtain the analytical expressions for the coefficients $q_E$ and $m$ for $Q_E$ and $\mathcal{M}$, respectively.
First, we consider the simulation accuracy $Q_E$ and afterwards the Trotter errors on the magnetization $\mathcal{M}$.
For the derivation of the corrections appearing in $Q_E$, we can use that the energy of the target Hamiltonian $H$ and therefore the simulation accuracy $Q_E$ exhibit a substantial overlap with the emergent conserved quantity $H_F$:
\be
	\langle  H_F (n\tau) \rangle_\tau = \langle H_F \rangle = \mathrm{const.}
	\label{eq:HF_constant_of_motion}
\ee
Here, $\langle  \mathcal{O} (n\tau) \rangle_\tau = \langle \psi_0 | e^{iH_F n\tau} \mathcal{O} e^{-iH_F n\tau} | \psi_0 \rangle$ denotes the full Trotterized time evolution with Trotter step size $\tau$ as in the main text.
Moreover, we define the expectation values in the initial state via $\langle \mathcal{O} \rangle = \langle \psi_0 | \mathcal{O} | \psi_0 \rangle$ and under the ideal time evolution as $\langle \mathcal{O}(t) \rangle = \langle \mathcal{O}(t) \rangle_{\tau=0}$.

In order to obtain all corrections to the desired order, we first have to express $H_F$ using the Magnus expansion up to second order in the Trotter step size, 
\be
	H_F = H + \tau \mathcal{C}_1 + \tau^2 \mathcal{C}_2 + \mathcal{O}(\tau^3) \, ,
\ee
with
\be
	\mathcal{C}_1 =\frac{i}{2} [H_X,H_Z],\quad \mathcal{C}_2 = -\frac{1}{12} [H_X-H_Z,[H_X,H_Z]] \, .
\ee
For convenience, we restrict the presentation from now on to a sequence of two elementary gates within one period, as we have for the case of the simulated quantum Ising chain. 
Using the above expansion for $H_F$ in combination with the conservation of $H_F$, one obtains for the energy deviation 
\begin{align}
	\Delta E(n\tau) = & \langle H(n\tau)\rangle_\tau - \langle H \rangle \, ,\\
	= & \tau \Delta \mathcal{C}_1(n\tau) + \tau^2 \Delta \mathcal{C}_2(n\tau) \, . \nonumber
\end{align}
where
\be
	\Delta \mathcal{C}_{\nu}(n\tau) =\langle \mathcal{C}_{\nu}\rangle - \langle \mathcal{C}_{\nu} (n\tau)\rangle_\tau, \quad \nu = 1,2 \, .
\ee
As a next step, we use time-dependent perturbation theory to determine the leading order in $\tau$ corrections of $\Delta \mathcal{C}_\nu(n\tau)$.
For this purpose, we write
\be
	e^{-iH_F t} = e^{-iHt} \, W(t), \quad W(t) = \mathcal{T} e^{-i \int_0^t dt' \, V(t')} \, ,
\ee
with $\mathcal{T}$ denoting the time-ordering prescription and
\be
	V(t) = e^{iHt}V e^{-iHt},\quad V = \tau \mathcal{C}_1 + \tau^2 \mathcal{C}_2 \, .
\ee
For the corrections to $\Delta E(n\tau)$ quadratic in $\tau$, we need to perform time-dependent perturbation theory to first order in $\tau$ for $\mathcal{C}_1$ and can neglect any $\tau$-dependent contributions for $\mathcal{C}_2$.

Let us first consider $\Delta \mathcal{C}_1(n\tau)$, which gives 
\be
	\Delta \mathcal{C}_1(n\tau) = \langle \mathcal{C}_1 \rangle - \langle \mathcal{C}_1(n\tau) \rangle -i\tau \int_0^{n\tau} dt' \, \langle [\mathcal{C}_1(t'),\mathcal{C}_1(n\tau)] \rangle \, .
\ee 
The time integral can be conveniently evaluated by recognizing that
\be
	\mathcal{C}_1 = \frac{i}{2} [H_X,H_Z] = \frac{i}{2} [H,H_Z] \, , 
\ee
since $H=H_X+H_Z$, and thus
\be
	\mathcal{C}_1(t) = \frac{1}{2} \frac{d}{dt} H_Z(t) \, .
\ee
This gives
\be
	\Delta \mathcal{C}_1(n\tau) = \langle \mathcal{C}_1 \rangle - \langle \mathcal{C}_1(n\tau) \rangle -\frac{i\tau}{2} \langle [H_Z(n\tau)-H_Z,\mathcal{C}_1(n\tau)] \rangle \, .
\ee
In the limit of $n\to \infty$, we can use the general property that expectation values of operators are governed by the so called diagonal ensemble~\cite{ETHReview2016}
\be
	\langle \mathcal{O}(n\tau) \rangle \stackrel{n\to\infty}{\longrightarrow} \sum_{\lambda} p_\lambda \langle \lambda | \mathcal{O} | \lambda \rangle \, ,
\ee
where $p_\lambda = | \langle \lambda | \psi_0\rangle |^2$ and $|\lambda\rangle$ is a full set of eigenstates for the target Hamiltonian $H$.
Using particular properties of the considered protocol, the above result for $\Delta \mathcal{C}_1(n\tau)$ can be simplified considerably.
We can use, for example, that $\langle \mathcal{C}_1 \rangle = 0$ and $\langle [H_Z,\mathcal{C}_1(n\tau)]\rangle=0$, because $|\psi_0\rangle$ is an eigenstate for $H_Z$, which finally yields
\be
	\Delta \mathcal{C}_1(n\tau) \stackrel{n\to\infty}{\longrightarrow} -\frac{\tau}{4} \sum_{\lambda} p_\lambda \langle \lambda | [H_Z,[H_Z,H_X]] | \lambda \rangle \, .
\ee
For the contributions to $\Delta E(n\tau)$ that are second order in $\tau$ stemming from $\Delta \mathcal{C}_2(n\tau)$, we can restrict to the zeroth order in time-dependent perturbation theory for $\langle \mathcal{C}_2(n\tau)\rangle_{\tau} $, i.e., we can replace $\langle \mathcal{C}_2(n\tau)\rangle_{\tau} \to \langle \mathcal{C}_2(n\tau)\rangle$.
This yields
\be
	\Delta \mathcal{C}_2(n\tau) \stackrel{n\to\infty}{\longrightarrow} \langle \mathcal{C}_2 \rangle - \sum_\lambda p_\lambda \langle \lambda | \mathcal{C}_2 | \lambda \rangle \, .
\ee
Collecting all contributions, we finally obtain
\begin{align}
	Q_E = \frac{\Delta E}{E_{T=\infty}-E_0} = q_E (h\tau)^2 +  \mathcal{O}[(h\tau)^3]
\end{align}
with
\begin{align}
	q_E = & \frac{1}{J^2 E_0} \left[ \langle \mathcal{C}_2 \rangle -  \sum_\lambda p_\lambda \langle \lambda | \mathcal{C}_2 | \lambda \rangle - \right. \nonumber \\
	& \left. - \frac{1}{4} \sum_{\lambda} p_\lambda \langle \lambda | [H_Z,[H_Z,H_X]] | \lambda \rangle \right] \, ,
\end{align}
where we have used that $E_{T=\infty}=0$.
This expression can be evaluated using full diagonalization, which provides access to all eigenstates $|\lambda\rangle$.
For the considered parameters of our simulations, we find $q_E = 0.18$, which is consistent with the full dynamical calculation in the small Trotter step limit, see Fig.~\ref{fig:figRobustness}d.

For estimating the lowest-order corrections in $\tau$ for other observables such as the magnetization $\mathcal{M}$, we cannot make direct use of the emergent conserved quantity $H_F$ as we could for the energy of the target Hamiltonian.
Still, we can perform time-dependent perturbation theory, which we now have to carry out up to second order.
Following the same steps as before, we obtain for the magnetization the following expression
\begin{align}
	\Delta \mathcal{M}(n\tau) & = \langle \mathcal{M}(n\tau) \rangle_\tau - \langle \mathcal{M}(n\tau) \rangle = \nonumber \\
	& =  \frac{\tau^2}{12} \big[ \langle \{ H_Z^2(n\tau),\mathcal{M}(n\tau)\} \rangle - E_Z^2 \langle \mathcal{M}(t) \rangle  \big] \nonumber \\
	& + i \frac{\tau^2}{6} \langle [\mathcal{C}_1(n\tau)-\mathcal{C}_1,\mathcal{M}(n\tau)] \rangle \nonumber \\
	& - \frac{5\tau^2}{12} \int_0^{n\tau} dt \langle \mathcal{C}_1(t) H_Z(t) M(n\tau) + \mathrm{h.c.} \rangle \, .
\end{align}
Here, $\{A,B\}=AB+BA$ denotes the anticommutator and $E_Z$ is given by $H_Z |\psi_0\rangle = E_Z | \psi_0\rangle$.
In the limit $n\to \infty$, we can again use that expectation values can be evaluated in the diagonal ensemble.
In addition, the expression involving the time integral can be formally solved by using the Lehman representation.
Finally, we obtain 
\be
	\Delta \mathcal{M}(n\tau) \stackrel{n\to\infty}{\longrightarrow} m (h\tau)^2 + \mathcal{O}[(h\tau)^3] \, ,
\ee
with
\begin{align}
	m &  = \frac{1}{12 J^2} \sum_{\lambda} p_\lambda \langle \lambda |  \{ H_Z^2,\mathcal{M}\} - E_Z^2 \mathcal{M}| \lambda \rangle \nonumber \\
	&  -\frac{1}{6 J^2} \sum_{\lambda} p_\lambda \mathrm{Re} \big [ \langle \lambda | [H_X,M] H_Z | \lambda\rangle \big] \nonumber \\
	& + \frac{1}{6J^2} \sum_{\lambda,\lambda'} \frac{p_\lambda}{E_\lambda-E_{\lambda'}} \mathrm{Re} \Big[ \langle \lambda |  [H_Z,H_X]H_Z | \lambda'\rangle \langle \lambda' | \mathcal{M} | \lambda \rangle \Big] \nonumber \\
	& +\frac{1}{6J^2} \sum_{\lambda,\lambda'} \frac{\langle \lambda | \mathcal{M} | \lambda \rangle}{E_\lambda-E_{\lambda'}} \mathrm{Re} \Big[ C_\lambda C_{\lambda'}^\ast \langle \lambda |  [H_Z,H_X]H_Z | \lambda'\rangle  \Big] \, ,
\end{align}
where $C_\lambda = \langle \lambda |\psi_0\rangle$ and $E_\lambda$ denotes the eigenenergies of the target Hamiltonian $H$ corresponding to the eigenstate $|\lambda\rangle$.
Using full diagonalization, we can again evaluate this expression yielding for our model a value of $m=0.05$, which we have used in Fig.~\ref{fig:figRobustness}b for the asymptotic small $\tau$ prediction and which matches well the result from the full dynamics.


\bibliographystyle{apsrev}
\bibliography{references}

\begin{thebibliography}{40}
\expandafter\ifx\csname natexlab\endcsname\relax\def\natexlab#1{#1}\fi
\expandafter\ifx\csname bibnamefont\endcsname\relax
  \def\bibnamefont#1{#1}\fi
\expandafter\ifx\csname bibfnamefont\endcsname\relax
  \def\bibfnamefont#1{#1}\fi
\expandafter\ifx\csname citenamefont\endcsname\relax
  \def\citenamefont#1{#1}\fi
\expandafter\ifx\csname url\endcsname\relax
  \def\url#1{\texttt{#1}}\fi
\expandafter\ifx\csname urlprefix\endcsname\relax\def\urlprefix{URL }\fi
\providecommand{\bibinfo}[2]{#2}
\providecommand{\eprint}[2][]{\url{#2}}

\bibitem[{\citenamefont{{Ladd T. D.} et~al.}(2010)\citenamefont{{Ladd T. D.},
  {Jelezko F.}, {Laflamme R.}, {Nakamura Y.}, {Monroe C.}, and {O Brien J.
  L.}}}]{Ladd2000}
\bibinfo{author}{\bibnamefont{{Ladd T. D.}}},
  \bibinfo{author}{\bibnamefont{{Jelezko F.}}},
  \bibinfo{author}{\bibnamefont{{Laflamme R.}}},
  \bibinfo{author}{\bibnamefont{{Nakamura Y.}}},
  \bibinfo{author}{\bibnamefont{{Monroe C.}}}, \bibnamefont{and}
  \bibinfo{author}{\bibnamefont{{O Brien J. L.}}}, \bibinfo{journal}{Nature}
  \textbf{\bibinfo{volume}{464}}, \bibinfo{pages}{45} (\bibinfo{year}{2010}).

\bibitem[{\citenamefont{Feynman}(1982)}]{Feynman1982}
\bibinfo{author}{\bibfnamefont{R.~P.} \bibnamefont{Feynman}},
  \bibinfo{journal}{Int. J. Theor. Phys.} \textbf{\bibinfo{volume}{21}},
  \bibinfo{pages}{467} (\bibinfo{year}{1982}).

\bibitem[{\citenamefont{Lloyd}(1996)}]{Lloyd1996}
\bibinfo{author}{\bibfnamefont{S.}~\bibnamefont{Lloyd}},
  \bibinfo{journal}{Science} \textbf{\bibinfo{volume}{273}},
  \bibinfo{pages}{1073} (\bibinfo{year}{1996}).

\bibitem[{\citenamefont{O'Malley et~al.}(2016)\citenamefont{O'Malley, Babbush,
  Kivlichan, Romero, McClean, Barends, Kelly, Roushan, Tranter, Ding
  et~al.}}]{OMalley2016}
\bibinfo{author}{\bibfnamefont{P.~J.~J.} \bibnamefont{O'Malley}},
  \bibinfo{author}{\bibfnamefont{R.}~\bibnamefont{Babbush}},
  \bibinfo{author}{\bibfnamefont{I.~D.} \bibnamefont{Kivlichan}},
  \bibinfo{author}{\bibfnamefont{J.}~\bibnamefont{Romero}},
  \bibinfo{author}{\bibfnamefont{J.~R.} \bibnamefont{McClean}},
  \bibinfo{author}{\bibfnamefont{R.}~\bibnamefont{Barends}},
  \bibinfo{author}{\bibfnamefont{J.}~\bibnamefont{Kelly}},
  \bibinfo{author}{\bibfnamefont{P.}~\bibnamefont{Roushan}},
  \bibinfo{author}{\bibfnamefont{A.}~\bibnamefont{Tranter}},
  \bibinfo{author}{\bibfnamefont{N.}~\bibnamefont{Ding}}, \bibnamefont{et~al.},
  \bibinfo{journal}{Phys. Rev. X} \textbf{\bibinfo{volume}{6}},
  \bibinfo{pages}{031007} (\bibinfo{year}{2016}).

\bibitem[{\citenamefont{Kandala et~al.}(2017)\citenamefont{Kandala, Mezzacapo,
  Temme, Takita, Brink, Chow, and Gambetta}}]{Kandala2017}
\bibinfo{author}{\bibfnamefont{A.}~\bibnamefont{Kandala}},
  \bibinfo{author}{\bibfnamefont{A.}~\bibnamefont{Mezzacapo}},
  \bibinfo{author}{\bibfnamefont{K.}~\bibnamefont{Temme}},
  \bibinfo{author}{\bibfnamefont{M.}~\bibnamefont{Takita}},
  \bibinfo{author}{\bibfnamefont{M.}~\bibnamefont{Brink}},
  \bibinfo{author}{\bibfnamefont{J.~M.} \bibnamefont{Chow}}, \bibnamefont{and}
  \bibinfo{author}{\bibfnamefont{J.~M.} \bibnamefont{Gambetta}},
  \bibinfo{journal}{Nature} \textbf{\bibinfo{volume}{549}},
  \bibinfo{pages}{242} (\bibinfo{year}{2017}).

\bibitem[{\citenamefont{Hempel et~al.}(2018)\citenamefont{Hempel, Maier,
  Romero, McClean, Monz, Shen, Jurcevic, Lanyon, Love, Babbush
  et~al.}}]{hempel2018quantum}
\bibinfo{author}{\bibfnamefont{C.}~\bibnamefont{Hempel}},
  \bibinfo{author}{\bibfnamefont{C.}~\bibnamefont{Maier}},
  \bibinfo{author}{\bibfnamefont{J.}~\bibnamefont{Romero}},
  \bibinfo{author}{\bibfnamefont{J.}~\bibnamefont{McClean}},
  \bibinfo{author}{\bibfnamefont{T.}~\bibnamefont{Monz}},
  \bibinfo{author}{\bibfnamefont{H.}~\bibnamefont{Shen}},
  \bibinfo{author}{\bibfnamefont{P.}~\bibnamefont{Jurcevic}},
  \bibinfo{author}{\bibfnamefont{B.}~\bibnamefont{Lanyon}},
  \bibinfo{author}{\bibfnamefont{P.}~\bibnamefont{Love}},
  \bibinfo{author}{\bibfnamefont{R.}~\bibnamefont{Babbush}},
  \bibnamefont{et~al.}, \bibinfo{journal}{arXiv:1803.10238}
  (\bibinfo{year}{2018}).

\bibitem[{\citenamefont{Lanyon et~al.}(2011)\citenamefont{Lanyon, Hempel, Nigg,
  M{\"u}ller, Gerritsma, Zaehringer, Schindler, Barreiro, Rambach, Kirchmair
  et~al.}}]{Lanyon2011}
\bibinfo{author}{\bibfnamefont{B.~P.} \bibnamefont{Lanyon}},
  \bibinfo{author}{\bibfnamefont{C.}~\bibnamefont{Hempel}},
  \bibinfo{author}{\bibfnamefont{D.}~\bibnamefont{Nigg}},
  \bibinfo{author}{\bibfnamefont{M.}~\bibnamefont{M{\"u}ller}},
  \bibinfo{author}{\bibfnamefont{R.}~\bibnamefont{Gerritsma}},
  \bibinfo{author}{\bibfnamefont{F.}~\bibnamefont{Zaehringer}},
  \bibinfo{author}{\bibfnamefont{P.}~\bibnamefont{Schindler}},
  \bibinfo{author}{\bibfnamefont{J.~T.} \bibnamefont{Barreiro}},
  \bibinfo{author}{\bibfnamefont{M.}~\bibnamefont{Rambach}},
  \bibinfo{author}{\bibfnamefont{G.}~\bibnamefont{Kirchmair}},
  \bibnamefont{et~al.}, \bibinfo{journal}{Science}
  \textbf{\bibinfo{volume}{334}}, \bibinfo{pages}{57} (\bibinfo{year}{2011}).

\bibitem[{\citenamefont{Salath{\'e} et~al.}(2015)\citenamefont{Salath{\'e},
  Mondal, Oppliger, Heinsoo, Kurpiers, Potocnik, Mezzacapo, Heras, Lamata,
  Solano et~al.}}]{Salathe2015}
\bibinfo{author}{\bibfnamefont{Y.}~\bibnamefont{Salath{\'e}}},
  \bibinfo{author}{\bibfnamefont{M.}~\bibnamefont{Mondal}},
  \bibinfo{author}{\bibfnamefont{M.}~\bibnamefont{Oppliger}},
  \bibinfo{author}{\bibfnamefont{J.}~\bibnamefont{Heinsoo}},
  \bibinfo{author}{\bibfnamefont{P.}~\bibnamefont{Kurpiers}},
  \bibinfo{author}{\bibfnamefont{A.}~\bibnamefont{Potocnik}},
  \bibinfo{author}{\bibfnamefont{A.}~\bibnamefont{Mezzacapo}},
  \bibinfo{author}{\bibfnamefont{U.~L.} \bibnamefont{Heras}},
  \bibinfo{author}{\bibfnamefont{L.}~\bibnamefont{Lamata}},
  \bibinfo{author}{\bibfnamefont{E.}~\bibnamefont{Solano}},
  \bibnamefont{et~al.}, \bibinfo{journal}{Phys. Rev. X}
  \textbf{\bibinfo{volume}{5}}, \bibinfo{pages}{021027} (\bibinfo{year}{2015}).

\bibitem[{\citenamefont{Barreiro et~al.}(2011)\citenamefont{Barreiro,
  M{\"u}ller, Schindler, Nigg, Monz, Chwalla, Hennrich, Roos, Zoller, and
  Blatt}}]{Barreiro2011}
\bibinfo{author}{\bibfnamefont{J.}~\bibnamefont{Barreiro}},
  \bibinfo{author}{\bibfnamefont{M.}~\bibnamefont{M{\"u}ller}},
  \bibinfo{author}{\bibfnamefont{P.}~\bibnamefont{Schindler}},
  \bibinfo{author}{\bibfnamefont{D.}~\bibnamefont{Nigg}},
  \bibinfo{author}{\bibfnamefont{T.}~\bibnamefont{Monz}},
  \bibinfo{author}{\bibfnamefont{M.}~\bibnamefont{Chwalla}},
  \bibinfo{author}{\bibfnamefont{M.}~\bibnamefont{Hennrich}},
  \bibinfo{author}{\bibfnamefont{C.}~\bibnamefont{Roos}},
  \bibinfo{author}{\bibfnamefont{P.}~\bibnamefont{Zoller}}, \bibnamefont{and}
  \bibinfo{author}{\bibfnamefont{R.}~\bibnamefont{Blatt}},
  \bibinfo{journal}{Nature} \textbf{\bibinfo{volume}{470}},
  \bibinfo{pages}{486} (\bibinfo{year}{2011}).

\bibitem[{\citenamefont{Barends et~al.}(2015)\citenamefont{Barends, Lamata,
  Kelly, Garcia-Alvarez, Fowler, Megrant, Jeffrey, White, Sank, Mutus
  et~al.}}]{Barends2015}
\bibinfo{author}{\bibfnamefont{R.}~\bibnamefont{Barends}},
  \bibinfo{author}{\bibfnamefont{L.}~\bibnamefont{Lamata}},
  \bibinfo{author}{\bibfnamefont{J.}~\bibnamefont{Kelly}},
  \bibinfo{author}{\bibfnamefont{L.}~\bibnamefont{Garcia-Alvarez}},
  \bibinfo{author}{\bibfnamefont{A.~G.} \bibnamefont{Fowler}},
  \bibinfo{author}{\bibfnamefont{A.}~\bibnamefont{Megrant}},
  \bibinfo{author}{\bibfnamefont{E.}~\bibnamefont{Jeffrey}},
  \bibinfo{author}{\bibfnamefont{T.~C.} \bibnamefont{White}},
  \bibinfo{author}{\bibfnamefont{D.}~\bibnamefont{Sank}},
  \bibinfo{author}{\bibfnamefont{J.~Y.} \bibnamefont{Mutus}},
  \bibnamefont{et~al.}, \bibinfo{journal}{Nature Commun.}
  \textbf{\bibinfo{volume}{6}}, \bibinfo{pages}{7654} (\bibinfo{year}{2015}).

\bibitem[{\citenamefont{Langford et~al.}(2017)\citenamefont{Langford,
  Sagastizabal, Kounalakis, Dickel, Bruno, Luthi, Thoen, Endo, and
  DiCarlo}}]{Langford2016}
\bibinfo{author}{\bibfnamefont{N.~K.} \bibnamefont{Langford}},
  \bibinfo{author}{\bibfnamefont{R.}~\bibnamefont{Sagastizabal}},
  \bibinfo{author}{\bibfnamefont{M.}~\bibnamefont{Kounalakis}},
  \bibinfo{author}{\bibfnamefont{C.}~\bibnamefont{Dickel}},
  \bibinfo{author}{\bibfnamefont{A.}~\bibnamefont{Bruno}},
  \bibinfo{author}{\bibfnamefont{F.}~\bibnamefont{Luthi}},
  \bibinfo{author}{\bibfnamefont{D.~J.} \bibnamefont{Thoen}},
  \bibinfo{author}{\bibfnamefont{A.}~\bibnamefont{Endo}}, \bibnamefont{and}
  \bibinfo{author}{\bibfnamefont{L.}~\bibnamefont{DiCarlo}},
  \bibinfo{journal}{Nature Commun.} \textbf{\bibinfo{volume}{8}},
  \bibinfo{pages}{1715} (\bibinfo{year}{2017}).

\bibitem[{\citenamefont{Wei et~al.}(2018)\citenamefont{Wei, Ramanathan, and
  Cappellaro}}]{Wei2018}
\bibinfo{author}{\bibfnamefont{K.~X.} \bibnamefont{Wei}},
  \bibinfo{author}{\bibfnamefont{C.}~\bibnamefont{Ramanathan}},
  \bibnamefont{and}
  \bibinfo{author}{\bibfnamefont{P.}~\bibnamefont{Cappellaro}},
  \bibinfo{journal}{Phys. Rev. Lett.} \textbf{\bibinfo{volume}{120}},
  \bibinfo{pages}{070501} (\bibinfo{year}{2018}).

\bibitem[{\citenamefont{Martinez et~al.}(2016)\citenamefont{Martinez, Muschik,
  Schindler, Nigg, Erhard, Heyl, Hauke, Dalmonte, Monz, Zoller
  et~al.}}]{Martinez2016}
\bibinfo{author}{\bibfnamefont{E.~A.} \bibnamefont{Martinez}},
  \bibinfo{author}{\bibfnamefont{C.~A.} \bibnamefont{Muschik}},
  \bibinfo{author}{\bibfnamefont{P.}~\bibnamefont{Schindler}},
  \bibinfo{author}{\bibfnamefont{D.}~\bibnamefont{Nigg}},
  \bibinfo{author}{\bibfnamefont{A.}~\bibnamefont{Erhard}},
  \bibinfo{author}{\bibfnamefont{M.}~\bibnamefont{Heyl}},
  \bibinfo{author}{\bibfnamefont{P.}~\bibnamefont{Hauke}},
  \bibinfo{author}{\bibfnamefont{M.}~\bibnamefont{Dalmonte}},
  \bibinfo{author}{\bibfnamefont{T.}~\bibnamefont{Monz}},
  \bibinfo{author}{\bibfnamefont{P.}~\bibnamefont{Zoller}},
  \bibnamefont{et~al.}, \bibinfo{journal}{Nature}
  \textbf{\bibinfo{volume}{534}}, \bibinfo{pages}{516} (\bibinfo{year}{2016}).

\bibitem[{\citenamefont{Berry et~al.}(2007)\citenamefont{Berry, Ahokas, Cleve,
  and Sanders}}]{Berry2007}
\bibinfo{author}{\bibfnamefont{D.~W.} \bibnamefont{Berry}},
  \bibinfo{author}{\bibfnamefont{G.}~\bibnamefont{Ahokas}},
  \bibinfo{author}{\bibfnamefont{R.}~\bibnamefont{Cleve}}, \bibnamefont{and}
  \bibinfo{author}{\bibfnamefont{B.~C.} \bibnamefont{Sanders}},
  \bibinfo{journal}{Commun. Math. Phys.} \textbf{\bibinfo{volume}{270}},
  \bibinfo{pages}{359} (\bibinfo{year}{2007}).

\bibitem[{\citenamefont{Poulin et~al.}(2015)\citenamefont{Poulin, Hastings,
  Wecker, Wiebe, Doherty, and Troyer}}]{poulin2014trotter}
\bibinfo{author}{\bibfnamefont{D.}~\bibnamefont{Poulin}},
  \bibinfo{author}{\bibfnamefont{M.~B.} \bibnamefont{Hastings}},
  \bibinfo{author}{\bibfnamefont{D.}~\bibnamefont{Wecker}},
  \bibinfo{author}{\bibfnamefont{N.}~\bibnamefont{Wiebe}},
  \bibinfo{author}{\bibfnamefont{A.~C.} \bibnamefont{Doherty}},
  \bibnamefont{and} \bibinfo{author}{\bibfnamefont{M.}~\bibnamefont{Troyer}},
  \bibinfo{journal}{QIC} \textbf{\bibinfo{volume}{15}}, \bibinfo{pages}{361}
  (\bibinfo{year}{2015}).

\bibitem[{\citenamefont{Babbush et~al.}(2015)\citenamefont{Babbush, McClean,
  Wecker, Aspuru-Guzik, and Wiebe}}]{Babbush2015}
\bibinfo{author}{\bibfnamefont{R.}~\bibnamefont{Babbush}},
  \bibinfo{author}{\bibfnamefont{J.}~\bibnamefont{McClean}},
  \bibinfo{author}{\bibfnamefont{D.}~\bibnamefont{Wecker}},
  \bibinfo{author}{\bibfnamefont{A.}~\bibnamefont{Aspuru-Guzik}},
  \bibnamefont{and} \bibinfo{author}{\bibfnamefont{N.}~\bibnamefont{Wiebe}},
  \bibinfo{journal}{Phys. Rev. A} \textbf{\bibinfo{volume}{91}},
  \bibinfo{pages}{022311} (\bibinfo{year}{2015}).

\bibitem[{\citenamefont{Anderson}(1958)}]{Anderson1958}
\bibinfo{author}{\bibfnamefont{P.~W.} \bibnamefont{Anderson}},
  \bibinfo{journal}{Phys. Rev.} \textbf{\bibinfo{volume}{109}},
  \bibinfo{pages}{1492} (\bibinfo{year}{1958}).

\bibitem[{\citenamefont{Basko et~al.}(2006)\citenamefont{Basko, Aleiner, and
  Altshuler}}]{BASKO2006}
\bibinfo{author}{\bibfnamefont{D.}~\bibnamefont{Basko}},
  \bibinfo{author}{\bibfnamefont{I.}~\bibnamefont{Aleiner}}, \bibnamefont{and}
  \bibinfo{author}{\bibfnamefont{B.}~\bibnamefont{Altshuler}},
  \bibinfo{journal}{Ann. Phys. (NY)} \textbf{\bibinfo{volume}{321}},
  \bibinfo{pages}{1126} (\bibinfo{year}{2006}).

\bibitem[{\citenamefont{D'Alessio and Polkovnikov}(2013)}]{DAlessio2013}
\bibinfo{author}{\bibfnamefont{L.}~\bibnamefont{D'Alessio}} \bibnamefont{and}
  \bibinfo{author}{\bibfnamefont{A.}~\bibnamefont{Polkovnikov}},
  \bibinfo{journal}{Ann. Phys. (NY)} \textbf{\bibinfo{volume}{333}},
  \bibinfo{pages}{19} (\bibinfo{year}{2013}).

\bibitem[{\citenamefont{Lazarides et~al.}(2014)\citenamefont{Lazarides, Das,
  and Moessner}}]{Lazarides2014}
\bibinfo{author}{\bibfnamefont{A.}~\bibnamefont{Lazarides}},
  \bibinfo{author}{\bibfnamefont{A.}~\bibnamefont{Das}}, \bibnamefont{and}
  \bibinfo{author}{\bibfnamefont{R.}~\bibnamefont{Moessner}},
  \bibinfo{journal}{Phys. Rev. E} \textbf{\bibinfo{volume}{90}},
  \bibinfo{pages}{012110} (\bibinfo{year}{2014}).

\bibitem[{\citenamefont{Abanin et~al.}(2015)\citenamefont{Abanin, Roeck, and
  Huveneers}}]{Abanin2015}
\bibinfo{author}{\bibfnamefont{D.~A.} \bibnamefont{Abanin}},
  \bibinfo{author}{\bibfnamefont{W.~D.} \bibnamefont{Roeck}}, \bibnamefont{and}
  \bibinfo{author}{\bibfnamefont{F.}~\bibnamefont{Huveneers}},
  \bibinfo{journal}{Phys. Rev. Lett.} \textbf{\bibinfo{volume}{115}},
  \bibinfo{pages}{256803} (\bibinfo{year}{2015}).

\bibitem[{\citenamefont{Mori et~al.}(2016)\citenamefont{Mori, Kuwahara, and
  Saito}}]{Mori2016}
\bibinfo{author}{\bibfnamefont{T.}~\bibnamefont{Mori}},
  \bibinfo{author}{\bibfnamefont{T.}~\bibnamefont{Kuwahara}}, \bibnamefont{and}
  \bibinfo{author}{\bibfnamefont{K.}~\bibnamefont{Saito}},
  \bibinfo{journal}{Phys. Rev. Lett.} \textbf{\bibinfo{volume}{116}},
  \bibinfo{pages}{120401} (\bibinfo{year}{2016}).

\bibitem[{\citenamefont{Peng et~al.}(2005)\citenamefont{Peng, Du, and
  Suter}}]{Peng2005}
\bibinfo{author}{\bibfnamefont{X.}~\bibnamefont{Peng}},
  \bibinfo{author}{\bibfnamefont{J.}~\bibnamefont{Du}}, \bibnamefont{and}
  \bibinfo{author}{\bibfnamefont{D.}~\bibnamefont{Suter}},
  \bibinfo{journal}{Phys. Rev. A} \textbf{\bibinfo{volume}{71}},
  \bibinfo{pages}{012307} (\bibinfo{year}{2005}).

\bibitem[{\citenamefont{Barends et~al.}(2016)\citenamefont{Barends, Shabani,
  Lamata, Kelly, Mezzacapo, Heras, Babbush, Fowler, Campbell, Chen
  et~al.}}]{Barends2016}
\bibinfo{author}{\bibfnamefont{R.}~\bibnamefont{Barends}},
  \bibinfo{author}{\bibfnamefont{A.}~\bibnamefont{Shabani}},
  \bibinfo{author}{\bibfnamefont{L.}~\bibnamefont{Lamata}},
  \bibinfo{author}{\bibfnamefont{J.}~\bibnamefont{Kelly}},
  \bibinfo{author}{\bibfnamefont{A.}~\bibnamefont{Mezzacapo}},
  \bibinfo{author}{\bibfnamefont{U.~L.} \bibnamefont{Heras}},
  \bibinfo{author}{\bibfnamefont{R.}~\bibnamefont{Babbush}},
  \bibinfo{author}{\bibfnamefont{A.~G.} \bibnamefont{Fowler}},
  \bibinfo{author}{\bibfnamefont{B.}~\bibnamefont{Campbell}},
  \bibinfo{author}{\bibfnamefont{Y.}~\bibnamefont{Chen}}, \bibnamefont{et~al.},
  \bibinfo{journal}{Nature} \textbf{\bibinfo{volume}{534}},
  \bibinfo{pages}{222} (\bibinfo{year}{2016}).

\bibitem[{\citenamefont{Jurcevic et~al.}(2017)\citenamefont{Jurcevic, Shen,
  Hauke, Maier, Brydges, Hempel, Lanyon, Heyl, Blatt, and Roos}}]{Jurcevic2016}
\bibinfo{author}{\bibfnamefont{P.}~\bibnamefont{Jurcevic}},
  \bibinfo{author}{\bibfnamefont{H.}~\bibnamefont{Shen}},
  \bibinfo{author}{\bibfnamefont{P.}~\bibnamefont{Hauke}},
  \bibinfo{author}{\bibfnamefont{C.}~\bibnamefont{Maier}},
  \bibinfo{author}{\bibfnamefont{T.}~\bibnamefont{Brydges}},
  \bibinfo{author}{\bibfnamefont{C.}~\bibnamefont{Hempel}},
  \bibinfo{author}{\bibfnamefont{B.~P.} \bibnamefont{Lanyon}},
  \bibinfo{author}{\bibfnamefont{M.}~\bibnamefont{Heyl}},
  \bibinfo{author}{\bibfnamefont{R.}~\bibnamefont{Blatt}}, \bibnamefont{and}
  \bibinfo{author}{\bibfnamefont{C.~F.} \bibnamefont{Roos}},
  \bibinfo{journal}{Phys. Rev. Lett.} \textbf{\bibinfo{volume}{119}},
  \bibinfo{pages}{080501} (\bibinfo{year}{2017}).

\bibitem[{\citenamefont{Sieberer et~al.}(2018)}]{Sieberer2018}
\bibinfo{author}{\bibfnamefont{L.}~\bibnamefont{Sieberer}}
  \bibnamefont{et~al.}, \bibinfo{journal}{in preparation}
  (\bibinfo{year}{2018}).

\bibitem[{\citenamefont{Haake}(2010)}]{Haake2010}
\bibinfo{author}{\bibfnamefont{F.}~\bibnamefont{Haake}},
  \emph{\bibinfo{title}{{Quantum Signatures of Chaos}}}
  (\bibinfo{publisher}{Springer}, \bibinfo{year}{2010}).

\bibitem[{\citenamefont{Ullah and Porter}(1963)}]{Ullah1963}
\bibinfo{author}{\bibfnamefont{N.}~\bibnamefont{Ullah}} \bibnamefont{and}
  \bibinfo{author}{\bibfnamefont{C.~E.} \bibnamefont{Porter}},
  \bibinfo{journal}{Phys. Rev.} \textbf{\bibinfo{volume}{132}},
  \bibinfo{pages}{948} (\bibinfo{year}{1963}).

\bibitem[{\citenamefont{Maldacena et~al.}(2016)\citenamefont{Maldacena,
  Shenker, and Stanford}}]{Maldacena2016}
\bibinfo{author}{\bibfnamefont{J.}~\bibnamefont{Maldacena}},
  \bibinfo{author}{\bibfnamefont{S.~H.} \bibnamefont{Shenker}},
  \bibnamefont{and} \bibinfo{author}{\bibfnamefont{D.}~\bibnamefont{Stanford}},
  \bibinfo{journal}{J. High Energy Phys.} \textbf{\bibinfo{volume}{2016}},
  \bibinfo{pages}{106} (\bibinfo{year}{2016}).

\bibitem[{\citenamefont{Kukuljan et~al.}(2017)\citenamefont{Kukuljan,
  Grozdanov, and Prosen}}]{Kukuljan2017}
\bibinfo{author}{\bibfnamefont{I.}~\bibnamefont{Kukuljan}},
  \bibinfo{author}{\bibfnamefont{S.}~\bibnamefont{Grozdanov}},
  \bibnamefont{and} \bibinfo{author}{\bibfnamefont{T.}~\bibnamefont{Prosen}},
  \bibinfo{journal}{Phys. Rev. B} \textbf{\bibinfo{volume}{96}},
  \bibinfo{pages}{060301} (\bibinfo{year}{2017}).

\bibitem[{\citenamefont{D'Alessio et~al.}(2016)\citenamefont{D'Alessio, Kafri,
  Polkovnikov, and Rigol}}]{ETHReview2016}
\bibinfo{author}{\bibfnamefont{L.}~\bibnamefont{D'Alessio}},
  \bibinfo{author}{\bibfnamefont{Y.}~\bibnamefont{Kafri}},
  \bibinfo{author}{\bibfnamefont{A.}~\bibnamefont{Polkovnikov}},
  \bibnamefont{and} \bibinfo{author}{\bibfnamefont{M.}~\bibnamefont{Rigol}},
  \bibinfo{journal}{Adv. Phys.} \textbf{\bibinfo{volume}{65}},
  \bibinfo{pages}{239} (\bibinfo{year}{2016}).

\bibitem[{\citenamefont{Kuwahara et~al.}(2016)\citenamefont{Kuwahara, Mori, and
  Saito}}]{KUWAHARA2016}
\bibinfo{author}{\bibfnamefont{T.}~\bibnamefont{Kuwahara}},
  \bibinfo{author}{\bibfnamefont{T.}~\bibnamefont{Mori}}, \bibnamefont{and}
  \bibinfo{author}{\bibfnamefont{K.}~\bibnamefont{Saito}},
  \bibinfo{journal}{Ann. Phys. (NY)} \textbf{\bibinfo{volume}{367}},
  \bibinfo{pages}{96 } (\bibinfo{year}{2016}).

\bibitem[{\citenamefont{Abanin et~al.}(2017)\citenamefont{Abanin, De~Roeck, Ho,
  and Huveneers}}]{Abanin2017}
\bibinfo{author}{\bibfnamefont{D.~A.} \bibnamefont{Abanin}},
  \bibinfo{author}{\bibfnamefont{W.}~\bibnamefont{De~Roeck}},
  \bibinfo{author}{\bibfnamefont{W.~W.} \bibnamefont{Ho}}, \bibnamefont{and}
  \bibinfo{author}{\bibfnamefont{F.~m.~c.} \bibnamefont{Huveneers}},
  \bibinfo{journal}{Phys. Rev. B} \textbf{\bibinfo{volume}{95}},
  \bibinfo{pages}{014112} (\bibinfo{year}{2017}).

\bibitem[{\citenamefont{Bernien et~al.}(2017)\citenamefont{Bernien, Schwartz,
  Keesling, Levine, Omran, Pichler, Choi, Zibrov, Endres, Greiner
  et~al.}}]{bernien2017}
\bibinfo{author}{\bibfnamefont{H.}~\bibnamefont{Bernien}},
  \bibinfo{author}{\bibfnamefont{S.}~\bibnamefont{Schwartz}},
  \bibinfo{author}{\bibfnamefont{A.}~\bibnamefont{Keesling}},
  \bibinfo{author}{\bibfnamefont{H.}~\bibnamefont{Levine}},
  \bibinfo{author}{\bibfnamefont{A.}~\bibnamefont{Omran}},
  \bibinfo{author}{\bibfnamefont{H.}~\bibnamefont{Pichler}},
  \bibinfo{author}{\bibfnamefont{S.}~\bibnamefont{Choi}},
  \bibinfo{author}{\bibfnamefont{A.~S.} \bibnamefont{Zibrov}},
  \bibinfo{author}{\bibfnamefont{M.}~\bibnamefont{Endres}},
  \bibinfo{author}{\bibfnamefont{M.}~\bibnamefont{Greiner}},
  \bibnamefont{et~al.}, \bibinfo{journal}{Nature}
  \textbf{\bibinfo{volume}{551}}, \bibinfo{pages}{579} (\bibinfo{year}{2017}).

\bibitem[{\citenamefont{Zhang et~al.}(2017)\citenamefont{Zhang, Pagano, Hess,
  Kyprianidis, Becker, Kaplan, Gorshkov, Gong, and Monroe}}]{Zhang2017}
\bibinfo{author}{\bibfnamefont{J.}~\bibnamefont{Zhang}},
  \bibinfo{author}{\bibfnamefont{G.}~\bibnamefont{Pagano}},
  \bibinfo{author}{\bibfnamefont{P.~W.} \bibnamefont{Hess}},
  \bibinfo{author}{\bibfnamefont{A.}~\bibnamefont{Kyprianidis}},
  \bibinfo{author}{\bibfnamefont{P.}~\bibnamefont{Becker}},
  \bibinfo{author}{\bibfnamefont{H.}~\bibnamefont{Kaplan}},
  \bibinfo{author}{\bibfnamefont{A.~V.} \bibnamefont{Gorshkov}},
  \bibinfo{author}{\bibfnamefont{Z.-X.} \bibnamefont{Gong}}, \bibnamefont{and}
  \bibinfo{author}{\bibfnamefont{C.}~\bibnamefont{Monroe}},
  \bibinfo{journal}{Nature} \textbf{\bibinfo{volume}{551}},
  \bibinfo{pages}{601} (\bibinfo{year}{2017}).

\bibitem[{\citenamefont{Otterbach et~al.}(2017)\citenamefont{Otterbach,
  Manenti, Alidoust, Bestwick, Block, Bloom, Caldwell, Didier, Fried, Hong
  et~al.}}]{Rigetti2017}
\bibinfo{author}{\bibfnamefont{J.}~\bibnamefont{Otterbach}},
  \bibinfo{author}{\bibfnamefont{R.}~\bibnamefont{Manenti}},
  \bibinfo{author}{\bibfnamefont{N.}~\bibnamefont{Alidoust}},
  \bibinfo{author}{\bibfnamefont{A.}~\bibnamefont{Bestwick}},
  \bibinfo{author}{\bibfnamefont{M.}~\bibnamefont{Block}},
  \bibinfo{author}{\bibfnamefont{B.}~\bibnamefont{Bloom}},
  \bibinfo{author}{\bibfnamefont{S.}~\bibnamefont{Caldwell}},
  \bibinfo{author}{\bibfnamefont{N.}~\bibnamefont{Didier}},
  \bibinfo{author}{\bibfnamefont{E.~S.} \bibnamefont{Fried}},
  \bibinfo{author}{\bibfnamefont{S.}~\bibnamefont{Hong}}, \bibnamefont{et~al.},
  \bibinfo{journal}{arXiv preprint arXiv:1712.05771}  (\bibinfo{year}{2017}).

\bibitem[{\citenamefont{Neill et~al.}(2018)\citenamefont{Neill, Roushan,
  Kechedzhi, Boixo, Isakov, Smelyanskiy, Megrant, Chiaro, Dunsworth, Arya
  et~al.}}]{Neill2018}
\bibinfo{author}{\bibfnamefont{C.}~\bibnamefont{Neill}},
  \bibinfo{author}{\bibfnamefont{P.}~\bibnamefont{Roushan}},
  \bibinfo{author}{\bibfnamefont{K.}~\bibnamefont{Kechedzhi}},
  \bibinfo{author}{\bibfnamefont{S.}~\bibnamefont{Boixo}},
  \bibinfo{author}{\bibfnamefont{S.}~\bibnamefont{Isakov}},
  \bibinfo{author}{\bibfnamefont{V.}~\bibnamefont{Smelyanskiy}},
  \bibinfo{author}{\bibfnamefont{A.}~\bibnamefont{Megrant}},
  \bibinfo{author}{\bibfnamefont{B.}~\bibnamefont{Chiaro}},
  \bibinfo{author}{\bibfnamefont{A.}~\bibnamefont{Dunsworth}},
  \bibinfo{author}{\bibfnamefont{K.}~\bibnamefont{Arya}}, \bibnamefont{et~al.},
  \bibinfo{journal}{Science} \textbf{\bibinfo{volume}{360}},
  \bibinfo{pages}{195} (\bibinfo{year}{2018}).

\bibitem[{\citenamefont{Schindler et~al.}(2011)\citenamefont{Schindler,
  Barreiro, Monz, Nebendahl, Nigg, Chwalla, Hennrich, and
  Blatt}}]{Schindler2011}
\bibinfo{author}{\bibfnamefont{P.}~\bibnamefont{Schindler}},
  \bibinfo{author}{\bibfnamefont{J.~T.} \bibnamefont{Barreiro}},
  \bibinfo{author}{\bibfnamefont{T.}~\bibnamefont{Monz}},
  \bibinfo{author}{\bibfnamefont{V.}~\bibnamefont{Nebendahl}},
  \bibinfo{author}{\bibfnamefont{D.}~\bibnamefont{Nigg}},
  \bibinfo{author}{\bibfnamefont{M.}~\bibnamefont{Chwalla}},
  \bibinfo{author}{\bibfnamefont{M.}~\bibnamefont{Hennrich}}, \bibnamefont{and}
  \bibinfo{author}{\bibfnamefont{R.}~\bibnamefont{Blatt}},
  \bibinfo{journal}{Science} \textbf{\bibinfo{volume}{332}},
  \bibinfo{pages}{1059} (\bibinfo{year}{2011}).

\bibitem[{\citenamefont{Reed et~al.}(2012)\citenamefont{Reed, DiCarlo, Nigg,
  Sun, Frunzio, Girvin, and Schoelkopf}}]{Reed2012}
\bibinfo{author}{\bibfnamefont{M.~D.} \bibnamefont{Reed}},
  \bibinfo{author}{\bibfnamefont{L.}~\bibnamefont{DiCarlo}},
  \bibinfo{author}{\bibfnamefont{S.~E.} \bibnamefont{Nigg}},
  \bibinfo{author}{\bibfnamefont{L.}~\bibnamefont{Sun}},
  \bibinfo{author}{\bibfnamefont{L.}~\bibnamefont{Frunzio}},
  \bibinfo{author}{\bibfnamefont{S.~M.} \bibnamefont{Girvin}},
  \bibnamefont{and} \bibinfo{author}{\bibfnamefont{R.~J.}
  \bibnamefont{Schoelkopf}}, \bibinfo{journal}{Nature}
  \textbf{\bibinfo{volume}{482}}, \bibinfo{pages}{382} (\bibinfo{year}{2012}).

\bibitem[{\citenamefont{Gardiner and Zoller}(2000)}]{Gardiner2000}
\bibinfo{author}{\bibfnamefont{C.~W.} \bibnamefont{Gardiner}} \bibnamefont{and}
  \bibinfo{author}{\bibfnamefont{P.}~\bibnamefont{Zoller}},
  \emph{\bibinfo{title}{{Quantum Noise}}} (\bibinfo{publisher}{Springer},
  \bibinfo{address}{Germany}, \bibinfo{year}{2000}), \bibinfo{edition}{second
  enlarged} ed.

\end{thebibliography}




\renewcommand{\theequation}{S\arabic{equation}}
\setcounter{equation}{0}
\renewcommand{\thefigure}{S\arabic{figure}}
\setcounter{figure}{0}
\renewcommand{\thesection}{S\arabic{section}}
\setcounter{section}{0}

\onecolumngrid


\vspace*{2cm}
\begin{center}
	{\bf \Large 
		Supplementary Materials to\vspace*{0.3cm}\\ 
		\emph{Quantum localization bounds Trotter errors in digital quantum simulation}
	}
\end{center}

\vspace*{0.3cm}
{\center{
		\hspace*{0.1\columnwidth}\begin{minipage}[c]{0.8\columnwidth}
			In these Supplementary Materials, 
			(i)  we provide numerical data for a second benchmark model for ditigal quantum simulation (DQS), the lattice Schwinger model; 
			(ii) and we discuss the influence of two typical extrinsic sources for imperfections on digital quantum simulators.
		\end{minipage}
	}
}

\section{Trotter errors in the lattice Schwinger model}

In order to demonstrate the generality of our results, we provide in this Section of the Supplementary Materials an analysis of a second benchmark example -- the lattice Schwinger model of 1+1D quantum electrodynamics (QED), which has recently been realized in a DQS~\cite{Martinez2016}.
In order to map the lattice Schwinger model to a pure spin system, this model can be described by the Hamiltonian 
\be
\label{eq:HSM}
H_\mathrm{SM} = H_\pm + H_Z,
\ee
where
\be
H_\pm = \sum_{l=1}^{N-1} H_\pm^l, \quad H_\pm^l = \frac w 2  \left[ \sigma_l^x \sigma_{l+1}^x + \sigma_l^y \sigma_{l+1}^y \right],
\ee
with
\be
H_Z = \frac{m}{2} \sum_{l=1}^N (-1)^l \sigma_l^z + J \sum_{l=1}^N L_n^2, \quad L_n = \frac{1}{2} \sum_{l=1}^n [\sigma_l^z + (-1)^l].
\ee
Here, $m$ is the rest mass of the fermionic particles and anti-particles, and $w$ describes their kinetic energy. The term $\propto J$ is the energy of the U(1) gauge fields. Using the Gauss law, these have been integrated out at the cost of introducing asymmetric long-range interactions between the fermions. The model describes a full, interacting lattice gauge theory, and no general analytic or numeric method exists to exactly compute its real-time dynamics, except in limiting cases or for small systems. This makes it a relevant target for DQS. 

In our numerical simulations using exact diagonalization, we have choose the following gate sequence to mimic the DQS:
\be
U^{(1)} = U_1 U_2 U_3,
\ee
with
\be
U_1 = e^{-i\tau H_Z}, U_{2} = e^{-i\tau \sum_{l=1}^{N/2}H_\pm^{2l-1}}, U_{3} = e^{-i\tau \sum_{l=1}^{N/2}H_\pm^{2l}}.
\ee
Following the recent experiment~\cite{Martinez2016}, we initialize the system in the bare vacuum, which in the spin-$1/2$ language corresponds to a simple Neel state, 
\be
|\psi_0\rangle = |\uparrow \downarrow \dots \uparrow \downarrow \rangle.
\ee
As for the Ising model in the main text, we compute the dynamics of observables for $2\cdot 10^4$ periods numerically using a Lanczos algorithm with full reorthogonalization and extract the long-time value of the studied quantities by performing a stroboscopic mean over the last $10^4$ periods.
For all the shown data, we use a fixed parameter set  with $w/J=m/J=1$.

\begin{figure}
	\centering
	\includegraphics[width=0.5\columnwidth]{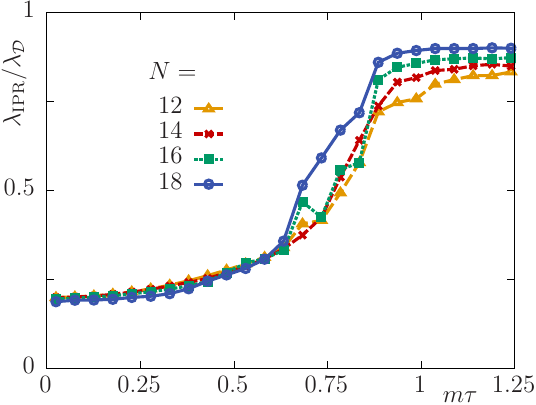}
	\caption{{\bf Inverse participation ratio for the DQS of the lattice Schwinger model.} The data is shown for different system sizes $N$. As for the Ising model, a sharp crossover divides a perturbative region at small $\tau$ from a fully quantum-chaotic regime at large $\tau$.}
	\label{fig:schwingerIPR}
\end{figure}

Again, we find a sharp quantum many-body chaos threshold in the inverse participation ratio (IPR), which we plot in Fig.~\ref{fig:schwingerIPR}.
For large Trotter steps $\tau$, we find that $\lambda_\mathrm{IPR} \to \lambda_\mathcal{D}$ implying full delocalization over all accessible states, whereas for small $\tau$ the system remains constrained and localized.
For $\lambda_\mathcal{D} = -N^{-1} \log(\mathcal{D})$, with $\mathcal{D}$ the number of accessible states in Hilbert space, we have incorporated the leading-order finite-size corrections as follows.
Random matrix theory predicts that  for a finite-sized system, the IPR is given by 
\be
\mathrm{IPR} = \frac{2}{\mathcal{D}} \, .
\ee
The lattice Schwinger model, however, exhibits an additional $U(1)$-symmetry---the conservation of the total spin---which is absent in the Ising chain considered in the main text. 
In the zero magnetization sector fixed by the initial condition, this implies that the total number of accessible states $\mathcal{D}$ is given by 
\be
\mathcal{D} = \frac{N!}{[(N/2)!]^2} \, ,
\ee
which is the value used for $\lambda_\mathcal{D}$ in Fig.~\ref{fig:schwingerIPR}.

\begin{figure}
	\centering
	\includegraphics[width=0.5\columnwidth]{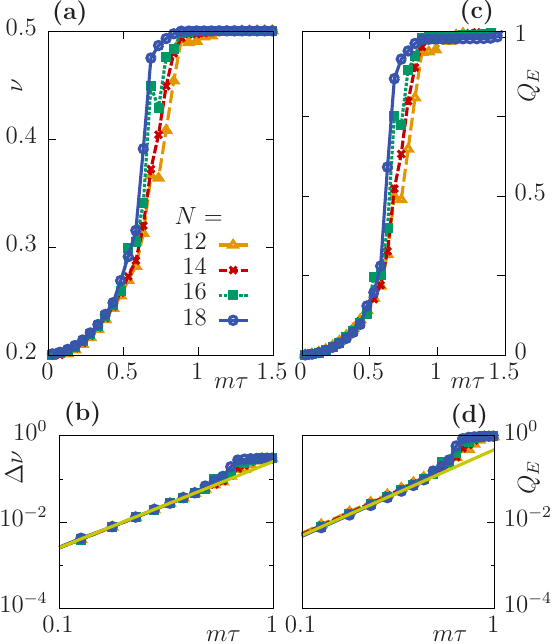}
	\caption{{\bf Trotter errors for the DQS of the lattice Schwinger model.} In {\bf (a)} we show the long-time value of the particle production $\nu$ and in {\bf (c)} the simulation accuracy $Q_E$. For small steps $\tau$ the Trotter error scales quadratically with $\tau$ for both quantities as shown for the deviation from the ideal result $\Delta\nu$ for the particle production in {\bf (b)} and the simulation accuracy $Q_E$ in {\bf (d)}.}
	\label{fig:schwingerObservables}
\end{figure}

Moreover, in Fig.~\ref{fig:schwingerObservables}a,b, we display the asymptotic long-time value $\nu$ of an important local observable of the lattice Schwinger model, the particle number relative to the bare vacuum 
\be
\nu(t)=\frac{1}{2N} \sum_l [(-1)^l \langle \sigma_l^z(t) \rangle + 1 ]. 
\ee
This quantity has also been measured in the recent experiment of Ref.~\cite{Martinez2016}. 
Complete scrambling corresponds to $\langle \sigma_l^z(t)\rangle \to 0$ for $t\to\infty$ and therefore $\nu(t) \to 1/2$.
In agreement with our results for the IPR in Fig.~\ref{fig:schwingerIPR}, the quantum many-body chaotic phase for large Trotter steps leads to uncontrolled Trotter errors.
In the localized phase on the other hand, the error becomes controllable, again with a quadratic dependence of the deviation $\Delta \nu$ from the ideal result on the Trotter step size $\tau$, see Fig.~\ref{fig:schwingerObservables}b.
As it can be seen in  Figs.~\ref{fig:schwingerObservables}c and ~\ref{fig:schwingerObservables}d, the simulation accuracy $Q_E$ signals the quantum many-body chaos threshold in a similar way.

\section{Imperfections\label{sec:imperfections}}

As mentioned in the main text, realistic experimental realizations of DQS not only face imperfections due to Trotterization but are also subject to various other error sources. In this Section, we address two generic error sources and discuss their implications. First, a timing error which results from inaccurate gate lengths, and, second, an ensemble error originating from slow drifts of the gate couplings. 

\subsection{Timing error}

Suppose that the gate length $\tau$ cannot be implemented perfectly but slightly fluctuates at each pulse. 
The gates performed at time step $p=1\dots n$ are then not the desired $\tau H_l$ but rather gates of slightly different strength, $\tau (1+\xi_l^p) H_l$, where $\xi_l^p$ are independent random variables with vanishing mean and variance $\propto\eta$. 

\begin{figure}
	\centering
	\includegraphics[width=0.5\columnwidth]{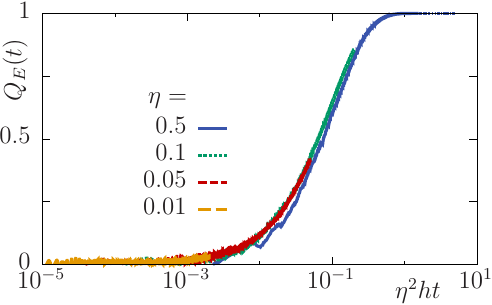}
	\caption{{\bf Timing errors in the dynamics of the simulation accuracy $Q_E(t)$ for the Ising model.} This data has been obtained for $N=18$ and varying noise strengths $\eta$ by averaging over $100$ noise realizations. The time axis has been rescaled with $\eta^2$ leading to a collapse of the data. As this shows, the time scale at which the timing error becomes relevant is proportional to $\eta^{-2}$.}
	\label{fig:errors}
\end{figure}

In Fig.~\ref{fig:errors}, we show an example numerical simulation for the Ising model used in the main text, with such an additional timing noise drawn from a uniform distribution of width $\eta$. When rescaling the time axis with $\eta^2$, we find a good collapse of the dynamics with slight deviations for large values of $\eta$. This finding implies that the accuracy of the DQS is not affected on a time scale proportional to $\eta^{-1}$ but rather on a much longer time scale proportional to $\eta^{-2}$.

This scaling behavior can be understood by mapping the timing errors to a Master equation. 
With timing errors, the effective Hamiltonian of Eq.~(3) of the main text depends on the time step $p$ and reads $\mathcal{H}^p=\sum_l H_l + i \frac{\tau}{2} \sum_{l>m}[H_l,H_m] + \sum_{l}\xi_l^p H_l+\mathcal{O}((\tau+\xi)^2)$. 
Assuming the fluctuations to be uncorrelated between time steps and gates, $\braket{\xi_l^p \xi_{l'}^{p'}}=\delta_{l,l'}\delta_{p,p'}\eta^2$, and extending the definitions to continuous time $\xi_l(t)=\xi_l^p$, $t\in[p,p+1)\tau$, the fluctuating gates can be described as noise with power spectrum 
\begin{eqnarray}
S(\omega)&=&\lim_{T\to\infty}\frac 1 T \int_{0}^{T}\ud t\int_{0}^{T}\ud t' \ue^{i\omega(t-t')}\braket{\xi_l(t) \xi_{l'}(t')}\nonumber\\
&=&\frac{2(1-\cos(\omega\tau))}{(\omega\tau)^2}\eta^2\tau\,.
\end{eqnarray}
If the relevant frequencies in the many-body system are small compared to $\tau^{-1}$, i.e., in the fast-driven regime that we are interested in, the power spectrum becomes flat, $S(\omega)=\eta^2\tau (1 + \mathcal{O}((\omega\tau)^2) )$, corresponding to white noise.  
In that regime, averaging over timing-error realizations, the time evolution of the system is to leading order described by an effective Markovian Master equation with Lindblad operators $H_l$, \cite{Gardiner2000}
\begin{align}
\dot{\rho}&=-i[\,H+\tau\sum_{l>m}[H_l,H_m]\,,\,\rho\,]\nonumber\\
&+\frac{\eta^2 \tau}{2} \sum_l \left(2 H_l\rho H_l - H_l^2 \rho -\rho H_l^2\right)\,.
\end{align}
The first term describes the controlled time evolution under the effective Hamiltonian discussed in the main text, consisting of the time-averaged Hamiltonian $H$ as well as the perturbation induced by Trotterization of strength $\sim \tau ||\sum_{l>m}[H_l,H_m]||$. The fluctuating gate strengths, instead, lead to a heating of the system with a rate $\sim \eta^2\tau ||\sum_{l}H_l^2||$, i.e., suppressed by an additional factor $\eta^2$.
Thus, for times $t\ll 1/( \tau \eta^{2})$ the influence of the timing error onto the dynamics is insignificant whereas for $t \gtrsim 1/( \tau \eta^{2})$ it becomes severe.

\subsection{Ensemble error}

While the timing error leads to fast fluctuations of gate strengths, gate errors may also be correlated over long times due to slowly drifting experimental parameters. In the limit of very slow drifts, the gate strength can be taken as constant within one experimental run but as randomly changing between runs. 
Such errors can be taken into account by sampling the time evolution over a family of Hamiltonians $\tilde{H}_l=H_l+\Delta_l H_l$, where $\Delta_l$ are independent random variables with vanishing mean that are assumed constant for each run. Each of these Hamiltonians will generate a time evolution under slightly modified gates $\tilde{U}_l=\exp(-i \tilde{H}_lt)$. 
As opposed to the timing error, this ensemble error, however, does not lead to heating, but only averages the resulting expectation values over a range of slightly different Hamiltonians. In particular in the perturbative regime, this error will be rather benign, except when working in hypersensitive regimes where observables do not behave smoothly, such as close to quantum phase transitions.

\end{document}